# Spin-orbit coupling in quantum gases

Victor Galitski[1,2] and Ian B. Spielman[1]

[1]Joint Quantum Institute, National Institute of Standards and Technology, and University of Maryland, Gaithersburg, Maryland, 20899, USA.
[2]Condensed Matter Theory Center, University of Maryland, College Park, Maryland, 20742, USA

Spin-orbit coupling links a particle's velocity to its quantum mechanical spin, and is essential in numerous condensed matter phenomena, including topological insulators and Majorana fermions. In solid-state materials, spin-orbit coupling originates from the movement of electrons in a crystal's intrinsic electric field, which is uniquely prescribed. In contrast, for ultracold atomic systems, the engineered "material parameters" are tuneable: a variety of synthetic spin-orbit couplings can be engineered on demand using laser fields. Here we outline the current experimental and theoretical status of spin-orbit coupling in ultracold atomic systems, discussing unique features that enable physics impossible in any other known setting.

A particle's spin is quantized. In contrast to the angular momentum of an ordinary, i.e. classical, spinning top which can take on any value, measurements of an electron's spin angular momentum (or just "spin") along some direction can result in only two discrete values: $\pm\hbar/2$, commonly referred to as spin-up or spin-down. This internal degree of freedom has no classical counterpart; in contrast, a quantum particle's velocity is directly analogous to a classical particle's velocity. It is therefore no surprise that spin is a cornerstone to a variety of deeply quantum materials like quantum magnets[1] and topological insulators[2]. Spin-orbit coupling (SOC) intimately unites a particle's spin with its momentum, bringing quantum mechanics to the forefront; in materials, this often increases the energy scale at which quantum effects are paramount.

The practical utility of any material is determined, not only by its intrinsic functional behavior, but also by the energy or temperature scale at which that behavior is present. For example, the quantum Hall effects – rare examples of truly quantum physics where the spin is largely irrelevant – are relegated to highly specialized laboratories because these phenomena manifest themselves only under extreme conditions – at liquid-helium temperatures and high magnetic fields[3,4]. The integer quantum Hall effect (IQHE) was the first observed topological insulator (TI), but it has a broken time-reversal symmetry. This is in contrast with a new class of topological insulators (see Box 1), which rely on SOC instead of magnetic fields for their quantum properties, and are expected to retain their quantumness up to room temperature[2].

As fascinating and unusual as the already existing topological world of spin-orbit-coupled systems is, all this physics is largely based on a non-interacting picture of independent electrons filling up a prescribed topological landscape. But there is clearly physics beyond this, as suggested by the fractional quantum Hall effect (FQHE) materials, where interactions between electrons yield phenomena qualitatively different from those encountered in IQHE. In FQHE systems, the charged excitations are essentially just fraction of an electron – with fractional charge – a new type of emergent excitation with no analogue elsewhere in physics. Furthermore, even non-Abelian excitations are possible: a system can be in one of many states of equal energy in which "non-Abelions," are at the same location, and differ only by the sequence of events that created them. At zero magnetic field, strong interactions and strong spin-orbit coupling can also give rise to fractionalization in TIs: the emergence of excitations that are fundamentally different from the constituent particles. We currently know little about these fractional topological insulators, but we do know they should exist and also expect them to be stable at a much larger range of parameters and experimental temperatures than the FQHE: perhaps even up to room temperature in solids.

It is ironic then, that we focus on the most fundamental behavior of spin-orbit coupled systems using ultracold atoms at nano-Kelvin temperatures. These nominally low temperatures are often a deception, as what matters is not an absolute temperature scale, but rather the temperature relative to other energy scales in the system (e.g., the Fermi energy), and from this perspective, ultracold atom systems are often not that cold[5]. However, ultracold atomic systems are among the simplest and most controllable of quantum many-body systems. While only one type of SOC has been experimentally realized to date, realistic theoretical proposals to create a range of SOCs abound, many of which have no counterpart in material systems[6,7,8,9,10]. The laser-coupling technique first experimentally implemented by our team[11,12] – now implemented in laboratories around the world – is well suited to realize topological states with one-dimensional atomic systems[13]. In sharp contrast to solid-state systems, in which we do not control or even know with certainty all details of the complicated material structure, ultracold atoms are remarkable in that most aspects of their environment can be engineered in the laboratory. Also, their tunable interactions and their single-particle potentials are both well characterized: the full atomic Hamiltonian *is* indeed known. This provides a level of control unprecedented in condensed matter and allows one to address basic physics questions at the intersection of material science and many-body theory. To study material systems, theorists create "spherical-cow" models of real materials, while in cold atom physics experimentalists actually make spherical cows.

Interactions – even the simple, contact interactions present in cold atom systems – enrich the physics of quantum systems by engendering new phases and phenomena. For example, when combined with SOC, the celebrated superfluid-Mott insulator transition[14,15], gives rise to numerous magnetic phenomena both in the insulating and superfluid phases[16,17]. Such interacting systems are often impossible to treat exactly with current theoretical techniques, but cold-atom experiments can directly realize these systems and shed light on the complicated and often exotic physics mediated by the strong interactions. Likewise, by asking basic questions such as how do strong interactions destroy TIs - or create them - we can understand the mechanisms underlying fractional TIs. These exotic quantum states have not yet been observed, but are present in realistic theoretical descriptions of ultracold atoms with SOC[18,19].

Ultracold atoms with synthetic spin-orbit couplings[8] can not only shed light on the outstanding problems of condensed matter physics, but also yield completely new phenomena with no analogue elsewhere in physics. A notable example of such a unique system is spin-orbit coupled bosons with just two spin states: a synthetic spin-1/2 system. The existence of such particles with real spin-1/2 is prohibited in fundamental physics due to Pauli's spin-statistics theorem, but synthetic symmetries – imposed by restricting the states available to the atoms – relax these constraints allowing bosons with pseudospin-1/2 to exist[20]. Spin-orbit coupling also results in a wide array of new many-body quantum states including a zoo of exotic quantum spin states in spin-orbit-coupled Mott insulators[21,16,17], unusual spin-orbit-coupled Bose-Einstein condensates, with a symmetry protected degenerate ground state[22], and perhaps even strongly correlated composite fermion phases analogous to the FQHE states in electron systems[18].

These are just a few examples of phenomena from a veritable treasure trove of exciting physics that is just waiting to be uncovered in this emerging and fast-developing field.

**Basics of spin-orbit coupling**

In any context, SOC requires symmetry breaking since the coupling strength is related to velocity as measured in a preferred reference frame (such as an electron's velocity with respect to its host crystalline lattice, or an atom's velocity with respect to a reference frame defined by its illuminating laser beams). Conventional SOC thus results from relativistic quantum mechanics, where the spin is a fundamental and inseparable component of electrons as described by the Dirac equation. In the non-relativistic limit, the Dirac equation reduces to the familiar Schrödinger equation, with relativistic corrections including an important term coupling the electron's spin to its momentum and to gradients of external potentials. This is the fundamental origin of SOC, which underlies both the $\mathbf{L} \cdot \mathbf{S}$ coupling familiar in atomic and molecular systems and all spin-orbit (SO) phenomena in solids. SOC can most simply be understood in terms of the familiar $-\boldsymbol{\mu} \cdot \mathbf{B}$ Zeeman interaction between a particle's magnetic moment $\boldsymbol{\mu}$ parallel to the spin, and a magnetic field $\mathbf{B}$ present in the frame moving with the particle.

SOC is most familiar in traditional atomic physics where it gives rise to atomic fine-structure splitting, and it is from this context that it acquires its name: a coupling between an electrons spin and its orbital angular momentum about the nucleus. The electric field produced by the charged nucleus gives rise to a magnetic field in the reference frame moving with an orbiting electron (along with an anomalous factor of two resulting from the electron's non-inertial trajectory encircling the atomic center of mass), leading to a momentum-dependent effective Zeeman energy.

In materials, the connection to a momentum-dependent Zeeman energy is particularly clear. For example, the Lorentz-invariant Maxwell's equations dictate that a static electric field $\mathbf{E} = E_0 \mathbf{e}_z$ in the lab frame (at rest) gives a magnetic field $\mathbf{B}_{SO} = (E_0 \hbar / mc^2)(k_x \mathbf{e}_y - k_y \mathbf{e}_x)$ in the frame of an object moving with momentum $\mathbf{k}$, where $c$ is the speed of light in vacuum and $m$ is the particle's mass. The resulting momentum-dependent Zeeman interaction $-\boldsymbol{\mu} \cdot \mathbf{B} \sim \sigma_x k_y - \sigma_y k_x$ is known as Rashba SOC[23]. This often arises from the built-in electric field in two-dimensional semiconductor heterostructures resulting from asymmetries of the confining potential[24], and is depicted in Fig. **1**. Figure **1d** plots a typical spin-orbit dispersion relation, where the minima for each spin state (red or blue) is displaced from zero; in the case of Rashba SOC, this dispersion is axially symmetric, meaning that this double-well structure is replicated for motion in any direction in the $\mathbf{e}_x$-$\mathbf{e}_y$ plane. Because of the momentum-dependent Zeeman interaction, the equilibrium alignment of a particle's magnetic moment depends on its velocity. Quantum mechanically, this implies that the quantum mechanical eigenstates are generally momentum-dependent superpositions of the initial $|\uparrow\rangle$ and $|\downarrow\rangle$ spin states.

In most condensed matter systems, electrons move in a crystal potential and when there is a potential gradient on the average, effective SO interactions arise. These usually originates from: a lack of mirror symmetry in two-dimensional systems leading to the Rashba SOC described above[23]; or a lack of inversion symmetry in bulk crystals leading to other forms of spin-orbit coupling such as the linear Dresselhaus SOC[25], described by a Zeeman interaction $-\boldsymbol{\mu} \cdot \mathbf{B} \sim \sigma_x k_x - \sigma_y k_y$ reminiscent to that of Rashba SOC.

SOC phenomena are ubiquitous in solids and have been known to exist since the early days of quantum mechanics and band theory. However, these phenomena and the field of spintronics[26] have moved to the forefront of condensed matter research only recently. This renewal of interest was stimulated by a number of exciting proposals for spintronic devices, whose functionality hinges on an electric field dependent coupling between the electron spin and its momentum. Apart from these potential useful applications, SO coupled systems turned out to display an amazing variety of fundamentally new and fascinating phenomena: spin-Hall effects[27,28], topological insulators[2], Majorana[13] and Weyl fermions[29], exotic spin textures in disordered systems[30], to name just a few.

The problem of synthesizing Majorana fermions stands out as perhaps the most active and exciting area of research combining both profound fundamental physics and a potential for applications. Indeed Kitaev noticed that a Majorana fermion, being a linear combination of a particle and a hole, should not couple much to external sources of noise and as such should be protected from its debilitating effects and decoherence[31]. Furthermore, when many such Majorana entities are put together, they form a non-Abelian network that is capable of encoding and processing topological quantum information that may be ideal for quantum computing applications[32]. Spin-orbit-coupled superconductors in a magnetic field can host Majorana fermions[33], and creating such topological fermionic superfluids in spin-orbit-coupled quantum gases appears to be within experimental reach, and perhaps cold atoms may become the first experimental platform to create and manipulate non-Abelian quantum matter.

**Synthetic SOC in cold atomic gases**

As we have seen, SOC links a particle's spin to its momentum and in conventional systems, it is a relativistic effect originating from electrons moving through a material's intrinsic electric field. This physical mechanism for creating SOC –requiring electric fields at the trillions of V/m level for significant SOC – is wildly inaccessible in the laboratory. Such fields exist inside atoms and materials, but not in laboratories. Instead, we engineer SOC in systems of ultracold atoms, using two photon Raman transitions – each driven by a pair of laser beams with wavelength $\lambda$ – that change the internal atomic "spin." Physically, this Raman process corresponds to the absorption of a single photon from one laser beam and its stimulated re-emission into the second. Each of these photons carries a tiny momentum with magnitude $p_R = h/\lambda$ called the photon recoil momentum ($h$ is Planck's constant). Conservation of momentum implies that the atom must acquire the difference of these two momenta (equal to $2p_R$ for counter-propagating laser beams). In most materials, the photon recoil is negligibly small; indeed, in conventional condensed matter systems, the "optical transitions" are described as having no momentum change. Ultracold atoms, however, are at such low temperatures that the momentum of even a single optical photon is quite large. Thus as first put forward by Higbie and Stamper-Kurn[34], Raman transitions can provide the required velocity dependent link between the spin and momentum: because the Raman lasers resonantly couple the spin states together, when an atom is moving, its Doppler shift effectively tunes the lasers away from resonance, altering the coupling in a velocity-dependent way. Remarkably, nearly all SOC phenomena present in solids can potentially be engineered with cold atoms (and some already have), but in contrast to solids where SOC is an intrinsic material property, synthetic SOC in cold atoms can be controlled at will. Furthermore, unlike the common electron, laser-dressed atoms with their pseudo-spins are not constrained by fundamental symmetries; this leads to a remarkably broad array of "synthetically-engineered" physical phenomena not encountered anywhere else in physics.

Figure **2** depicts the currently implemented technique for creating SOC in ultracold atoms[12,35,36,37,38]. The first step, shown in Fig. **2a**, is to select from the many available internal atomic states a pair of states which we will associate with pseudo-spins states $|\uparrow\rangle$ and $|\downarrow\rangle$ that together comprise the atomic "spin." Two counterpropagating laser beams, which here define the x axis, couple this selected pair of atomic states to the atoms' motion along $\mathbf{e}_x$. Reminiscent of the case for Rashba SOC shown in Fig. 1**d**, this coupling alters the atom's energy-momentum dispersion, although here only motion along the $x$ direction is affected (Fig. 2**c**). In the standard language, both Rashba and Dresselhaus SOC are present, and have equal magnitude, giving the effective Zeeman shift $-\boldsymbol{\mu}\cdot\mathbf{B} \sim -\sigma_y k_x$. In solids, this symmetric combination of the Rashba and Dresselhaus coupling goes by the name of "persistent spin-helix symmetry point," where it on one hand allows spin control via SOC, but on the other minimizes the undesirable effect of spin memory loss[30].

Since, SOCs effect on a single particle is equivalent to that of a momentum-dependent Zeeman magnetic field, the particle's dispersion relation (e.g., the familiar kinetic energy $mv^2/2 = p^2/2m$ for a free particle) is split into two sub-bands corresponding to two spin-split components, now behaving differently (measured in Fig. 2**c**). For the linear SOC on which we focus, the band splitting simply shifts the minimum of the dispersion relation by an amount depending on the particle's internal state and the laser coupling strength. This effect, depicted in Fig. 2**b**, was first measured indirectly in Ref. 12 where a BEC was prepared in a mixture of $|\uparrow\rangle$ and $|\downarrow\rangle$ in each of the two minima of the dispersion, and the momentum of the two spins was measured as a function of laser intensity. More recently, the full dispersion curve was measured spectroscopically[38], clearly revealing the spin-orbit coupled structure as a function of momentum (Fig. 2**c**).

A panoply of SOCs can be created with additional lasers linking together additional internal states. Figure 3 shows a realistic example where three internal atomic states can be coupled, producing a tunable combination of Rashba and Dresselhaus SOC[39]. In these cases, one of the three initial atomic states is shifted by a large energy, leaving behind two pseudo-spins comprising a two-level system[6]. A further extension can generate an exotic 3D analog to the Rashba SOC, which we call Weyl SOC, that cannot exist in materials[10] or to types of SOC with more than the usual two spin states[9].

**Many-body physics**

An example of a unique quantum phenomenon made possible in ultracold atomic systems is spin-orbit coupled Bose-Einstein condensates. The main ingredient of these exotic many-body states are laser-dressed bosons with states $|\uparrow\rangle$ and $|\downarrow\rangle$ that create a synthetic spin-1/2 system. Because the Pauli spin-statistics theorem prohibits the existence of bosons with real spin-1/2, this is a weird and interesting entity by itself, but when many such entities are brought together in a SO coupled system, the degree of weirdness further increases. As the temperature is lowered, the bosons tend to condense, but in contrast to the conventional BEC, where the zero-momentum state is the unique state with lowest energy (the ground state is non-degenerate), SO bosons can have energy-momentum dispersion with several lowest energy states (the ground state is degenerate). For example, for Rashba=Dresselhaus SOC (Fig. 2**c**) there are two such minima; for pure Rashba SOC there is a continuous ring of minima (Fig. 1**d**); for the Weyl-type spin-orbit coupling there is a sphere of minima[10]. This is in contrast with the more conventional case of spinor BECs that include two or more spin states, but do not alter the energy-momentum dispersion relation.

The bosons' indecisiveness about what state to condense into is partially resolved by their interactions, which limits which states have lowest energy. But unless the interactions break a "synthetic time-reversal" (Kramers) symmetry, some degeneracy must remain, leading to the possibility of exotic states. For example, repulsive bosons with a non-equal combination of Rashba and Dresselhaus SOC are predicted to condense into a strongly entangled many-body "cat" state, where the whole condensate simultaneously is in a superposition of states with equal and opposite momentum. Such many-body cat states have long been sought in various experiments, but have never been convincingly observed. The SO BECs, existing in a double-well "potential" in momentum space (e.g., Fig. 1**d**) are promising in this regard because robust arguments support the existence of many-body cats[22]: (i) the symmetry protection of the exact spin degeneracy from splitting and (ii) an argument based on the Heisenberg uncertainty relation, which suggests that in order for the repulsive bosons to stay as far as possible from each other in real space, they should be as close as possible in dual momentum space. An experimental realization of such a many-body cat state would be a major scientific development.

On the experimental front, there are already exciting developments, which include the first realization of an Abelian spin-orbit coupling (corresponding to the persistent spin helix symmetry point, where Rashba and Dresselhaus SOCs are identical, see supporting text box for a discussion of the connection to Abelian and non-Abelian gauge fields) and observation of a spin-orbit-coupled Bose-Einstein condensate with rubidium atoms[12,35,36]. Exactly as expected, the time-of-flight images of cold SO coupled bosons feature two peaks that correspond to left- and right-moving condensates flying apart in the opposite directions. They however do not represent a cat state (where all the atoms are either in the left-moving or all in the right-moving condensate), but rather are either in a "striped" state (where all of the atoms are in the same state, which involves both positive and negative momenta), or in a phase separated state of the right- and left-moving condensates in the Abelian SO system[12,40,41,42] see Fig. 2**b**.

Spin-orbit-coupled ultracold fermions are intriguing[8]: even the behavior of two interacting fermions is fundamentally altered with the addition of SOC. Without SOC and in one spatial dimension, any attraction between two fermions, no matter how weak, always gives rise to the formation of a molecule. In two dimensions, the resulting molecular pairing is suppressed but not absent, with an exponentially small binding energy; and in three dimensions there is a threshold below-which there is no molecular state. However, in systems with many fermions, many-body effects guarantee the formation of Cooper pairs in any dimension, as long as attraction is present. The crossover between a BEC of molecules to a Bardeen-Cooper-Schrieffer (BCS) condensate of pairs is a smooth transition between physics described in terms of simple "native" molecules to the truly many-body physics of Cooper pairs[43,44]. SOC provides a completely different avenue for enhancing the pairing between two fermions. The ground-state of the Rashba SOC Hamiltonian consists of a one-dimensional ring in momentum and that of Weyl SOC is a two-dimensional sphere. This reduces the effective dimensionality and thereby strongly enhances molecular pairing. This ensures that there is no threshold for molecular formation in such SO systems and that the BEC-BCS crossover is strongly modified[10,45,46,47]. The many body physics of the BCS side is greatly affected as well. The main difficulty in realizing topological fermionic superfluids is creating the unconventional pairing mechanism between the atoms[48,49]. Such topological pairing has been proven difficult to achieve using $p$-wave Feshbach resonances due to debilitating effects of three-body losses[50]. SOC also can create effective interactions: for example, in analogy to the $d$-wave interactions recently demonstrated between colliding BECs[51], stable *p*-wave interactions generated by synthetic spin-orbit are expected and pave the way to atomic topological superfluids[52,53,54]. Experimentally, SOC in atomic Fermi systems has been realized in two labs[37,38], where the basic physical phenomena at the single particle level were confirmed.

**Outlook**

Spin-orbit coupled cold atoms represent a fascinating and fast-developing area of research significantly overlapping with traditional condensed matter physics, but importantly containing completely new phenomena not realizable anywhere else in nature. The potential for new experimental and theoretical understanding abounds.

Spin-orbit-coupled Bose-Einstein condensates and degenerate Fermi gases have now been realized in a handful of laboratories: the experimental study of these systems is just beginning. The immediate outlook centers on implementing the full range of SOCs that currently exist only in theoretical proposals: to date just one form of SOC has been engineered in the lab. To realize the true promise of these systems, a central experimental task is to engineer SOCs that link spin to momentum in two and three dimensions (non-abelian, and without analog in material systems). An unfortunate reality of light induced gauge fields, as currently envisioned, is the presence off-resonant light scattering – spontaneous emission – that leads to atom-loss, heating, or both. In the alkali atoms, this heating cannot be fully mitigated by selecting different laser parameters (like wavelength), as a result, an important direction of future research is finding schemes, or selecting different atomic species, where this problem is mitigated or absent.

Another thrust of research with synthetic SO-coupled fermions is to realize topological insulating states in optical lattices. A recent breakthrough in condensed matter physics is the understanding that the quantum Hall states represent just a tip of the iceberg of a zoo of topological states. A complete classification of those has by now been achieved for fermion systems in thermodynamic limit[55,56]. This leads to fundamentally different classes of Hamiltonians. For example, no non-trivial insulators exist in three dimensions if time-reversal invariance is allowed to be broken, but the by-now famous Z2 classification exists otherwise[57]. There are nine symmetry classes in each spatial dimension, however not all of them have been realized in solids.

In materials, the symmetries are usually "non-negotiable" while in "synthetic" SO systems, the symmetries and lack thereof can be controlled at will, opening the possibility to create and control topological states, including topological phases that are not realizable in solids. The ability to tune synthetic couplings suggests that a larger class of non-Abelian gauge structures is within the immediate experimental reach. These structures do not have analogues, or even names in solid state physics, but are most appropriately characterized as SU(3)-spin-orbit couplings. These can be created by focusing on a three-level manifold of dressed states, as opposed to two-level manifold corresponding to spin-up and down states for the usual SOC. The general coupling of the three internal dressed degrees of freedom to particle motion can not be spanned by three spin matrices, but requires 3x3 Gell-Mann matrices[58], which form generators of the SU(3) group well studied in the context of elementary particle physics. The algebraic structure, geometry, and topology of this complicated group are much different from the familiar spin case, and these differences will have profound observable manifestations.

A completely different way to create such topological matter is related to the non-equilibrium physics of spin-orbit-coupled systems. It is easy to experimentally engineer dynamic synthetic SOC and gauge fields with a prescribed time-dependence, which gives the opportunity to realize interesting dynamic structures, such as Floquet topological insulators[59] and Floquet Majorana fermions[60].

We expect that the most exciting physics in atomic SOC systems will rely on interactions, and lie at the intersection of experiment and theory. What is the physics of spin-orbit coupled Mott insulators and the corresponding superfluid-to-insulator phase transition? What is the ground state of the Rashba bosons, which were recently argued to undergo a statistical transmutation into fermions. How is the BEC-BCS transition altered by SOC? Each of these questions can only be answered in a partnership between experiment and theory: the underlying physics is so intricate that the correct answer is difficult to anticipate without direct measurement, and the meaning of these measurements can be inexplicable without theoretical guidance.

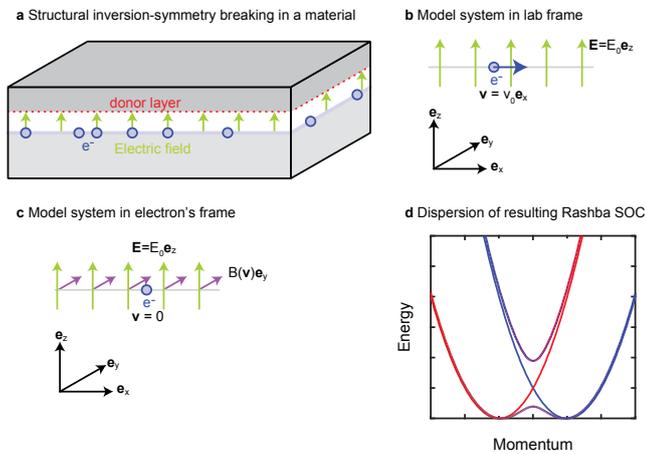

**Figure 1 | Physical origin of SOC in conventional systems. a** In materials, SOC requires a broken spatial symmetry. For example, the growth profile of two-dimensional GaAs electron (or hole) systems can create an intrinsic electric field breaking inversion symmetry. **b** The effective model system consists of an electron confined in the $\mathbf{e}_x$-$\mathbf{e}_y$ plane (in this example moving along $\mathbf{e}_x$) in the presence of an electric field along $\mathbf{e}_z$. **c** In the rest frame of the electron, the Lorentz-transformed electric field generates a magnetic field along $\mathbf{e}_y$ (generating a Zeeman shift ) which depends linearly on the electron's velocity. **d** For such systems the spin orbit coupling is linear, and the usual free particle $mv^2/2 = p^2/2m$ dispersion relation is altered in a spin dependent way. In this case, pure Rashba SOC shifts the free-particle dispersion relations for each spin state away from zero (red and blue curves). The crossing point of these curves can be split by an applied magnetic field (grey curve).

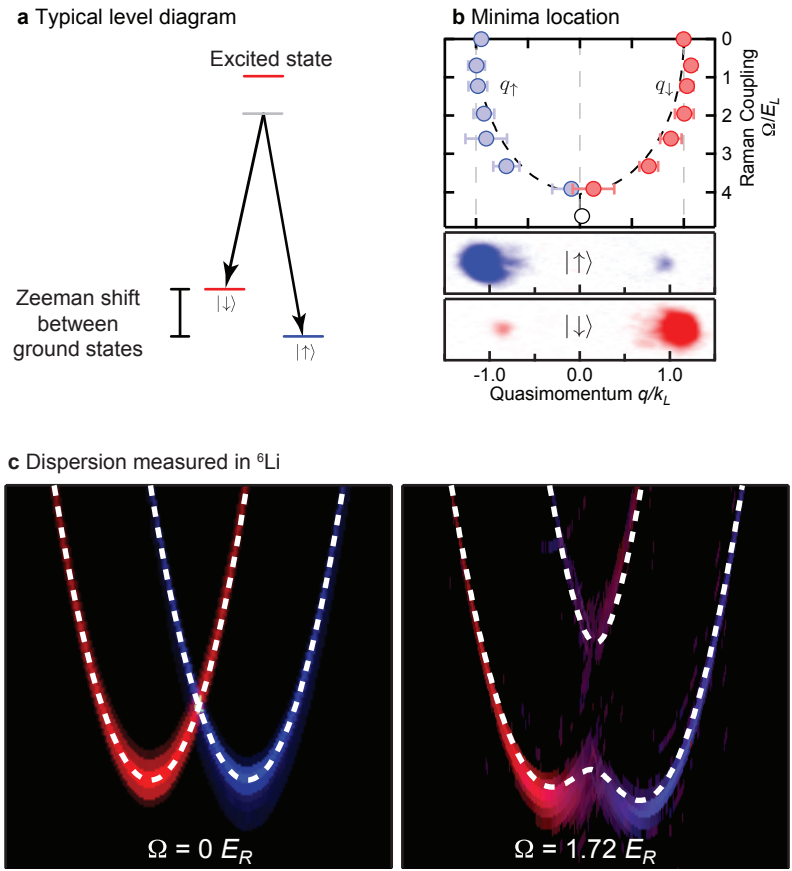

**Figure 2 | Laser coupling schemes. a.** In current experiments, a pair of lasers – often counter propagating – couple together a selected pair of atomic states labeled by $|\uparrow\rangle$ and $|\downarrow\rangle$ that together comprise the atomic "spin." These lasers are arranged in a two-photon Raman configuration that uses an off-resonant intermediate state (grey). These lasers link atomic motion along the *x* direction to the atom's spin creating a characteristic spin-orbit coupled energy-momentum dispersion relation. **b.** Measured location of energy minimum or minima, where as a function of laser intensity the characteristic double minima of SOC dispersion move together and finally merge[12]. **c.** Complete dispersion before and after laser coupling measured in a $^6$Li Fermi gas (Data reproduced with permission of M. Zwierlein, from Ref. 38), compared with the predicted dispersion (white dashed curves), showing typical spin-orbit dispersion relations depicted in Fig. **1d**.

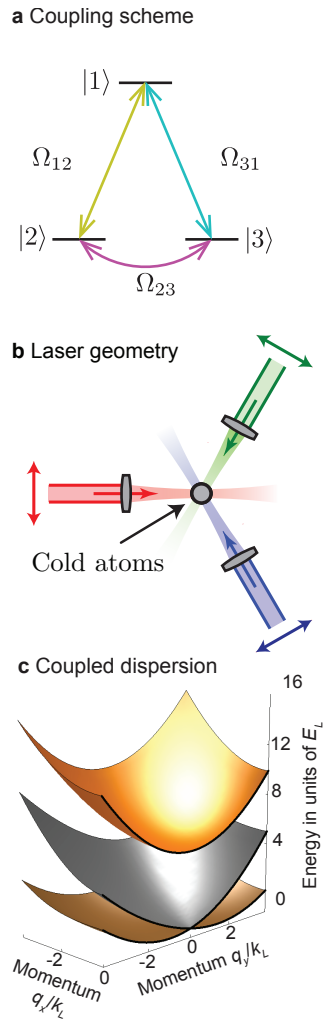

**Figure 3 | Generalized SOC.** **a-c.** Going beyond current experiments, more complicated forms of SOC may be created. These require both more laser beams and more internal states. **a** Each state is coupled by a two photon Raman transition, each produced by a pair of the beams shown in **b**. The depicted configuration could realize a tunable combination of Rashba and Dresselhaus SOC in the alkali atoms[39]; the outcome is equivalent to that of the well-known tripod configuration[6] with detuning, but practical in the alkali atoms.

**BOX 1**
# Topological matter

Topological insulators[2] are strongly spin-orbit-coupled materials that have seemingly mutually exclusive properties: they are both insulating and metallic at the same time. In their interior (bulk), electrons cannot propagate, while their surfaces are highly conducting. To get an insight into the complicated theory of these exotic materials, let us recall that electrons in an insulator fully occupy a certain number of allowed energies (bands) in such a way that the highest occupied state is separated from the lowest empty one by a gap of forbidden energies. Hence, a non-zero energy is required to excite an electron across the gap (i.e., to make it move) and small perturbations have almost no effect on the insulator. From this perspective, it is as good as vacuum: nothing moves inside. It may seem that any two such insulators ("vacua") should be indistinguishable, but it is not so! If we ask whether one insulator can be smoothly deformed into another without breaking certain symmetries or turning it into a metal along the way, we find that it is not always possible. Insulators are divided into qualitatively different categories, including trivial insulators (which are much like vacuum) and topological insulators, characterized by a non-zero integer topological index, related to the momentum-dependent spin in spin-orbit coupled materials such as occurs in HgTe/CdTe quantum wells[2] (preserving time-reversal symmetry) or from the magnetic field in quantum Hall systems (breaking time-reversal symmetry). Integers can not change smoothly one into another, but whenever we have a surface of a topological insulator, i.e., a boundary with a true vacuum, we effectively do enforce a transition between the media characterized by different integers, say 1 and 0, and the only way to cross between them is to either break symmetries or close the gap abruptly, that is to create a boundary metal. This is why the topological boundary states are so robust: they are squeezed in between the two vacua (the usual vacuum and the twisted one – the topological insulator) and have nowhere to go.

As different as superconductors are from insulators in their electromagnetic properties, the characterizations of their excitation spectra are closely related. A superconductor is a condensate of electron pairs (Cooper pairs) behaving like a superfluid. Since it is energetically favorable to form electron pairs in a superconductor, it takes energy to break a pair to create single electrons, just like it takes energy to move an electron across an energy gap in an insulator. So, a superconductor is an insulator for its fermionic excitations and as such can be characterized by topological integers with similar consequences, including boundary states. But the latter are unusual at the edges of a topological superconductor, which get filled by weird chargeless and spinless entities - linear combinations of an electron and a hole (an absentee electron). Under certain circumstances these can also become Majorana fermions – zero energy particles that are their own antiparticle – which were predicted in spin-orbit-coupled systems and might have been observed there[33].

**BOX 2**
# Connection to gauge fields

The forms of SOC discussed in this review are all examples of static gauge fields, which can be mathematically included in the atomic Hamiltonian as [       ]. The most elementary example of a gauge field is the electromagnetic vector potential defined by [     ], where $q$ is the electric charge of the particle. This vector potential defines magnetic and electric fields though its spatial and temporal properties; a uniform time-independent vector potential is of no physical consequence. A gauge field [ ] is non-Abelian when the components of the vector [       ] are non-commuting operators, for example [     ]. Such non-Abelian gauge potentials are generic in problems ranging from nuclear magnetic resonance to molecular collisions[61]. Using techniques related to those discussed here[62], it is possible to engineer artificial magnetic[11] and electric fields in ultracold atoms. In addition, two recent experiments have demonstrated two alternate techniques for creating artificial gauge fields: in the first, a spatially staggered magnetic field was generated using Raman-assisted tunneling in a 2D optical lattice[63,64]; and in the second, an artificial vector potential was created by carefully shaking a 1D optical lattice[65]. An intriguing direction for continued research is to create non-Abelian analogues to the electric and magnetic fields which result from variations in non-Abelian gauge fields[6,7], which would lead to quantum-gas analogs to the spin-Hall effect[66]. An exciting direction of research here is engineering dynamic gauge fields where the field is a dynamical quantum degree of freedom with analogs to quantum electro- and chromo-dynamics[67,68,69].

[1] Balents, L. Spin liquids in frustrated magnets. *Nature* **464**, 199–208 (2010).

[2] Hasan, M. Z. & Kane, C. L. Colloquium: Topological insulators. *Rev. Mod. Phys.* **82**, 3045–3067 (2010).

[3] Klitzing, von, K., Dorda, G. & Pepper, M. New method for high-accuracy determination of the fine-structure constant based on quantized Hall resistance. *Phys. Rev. Lett.* **45**, 494–497 (1980).

[4] Tsui, D. C., Stormer, H. L. & Gossard, A. C. Two-dimensional magnetotransport in the extreme quantum limit. *Phys. Rev. Lett.* **48**, 1559–1562 (1982).

[5] Bloch, I., Dalibard, J., and Zwerger, W. Many-body physics with ultracold gases. *Rev. Mod. Phys.* **80**, 885–964 (2008).

[6] Osterloh, K., Baig, M., Santos, L., Zoller, P. & Lewenstein, M. Cold Atoms in Non-Abelian Gauge Potentials: From the Hofstadter "Moth" to Lattice Gauge Theory. *Phys. Rev. Lett.* **95**, 010403 (2005). **An initial proposal suggesting a method to create SOC in cold atoms (equivalent to a non-abelian gauge field), in a lattice potential.**

[7] Ruseckas, J., Juzeliūnas, G., Ohberg, P. & Fleischhauer, M. Non-Abelian Gauge Potentials for Ultracold Atoms with Degenerate Dark States. *Phys. Rev. Lett.* **95**, 010404 (2005). **An initial proposal suggesting a method to create SOC in cold atoms (equivalent to a non-abelian gauge field) in the continuum.**

[8] X.-J. Liu, M. F. Borunda, X. Liu, and J. Sinova, Effect of Induced Spin-Orbit Coupling for Atoms via Laser Fields. *Phys. Rev. Lett.*, **102**, 046402 (2009).

[9] Juzeliūnas, G., Ruseckas, J. & Dalibard, J. Generalized Rashba-Dresselhaus spin-orbit coupling for cold atoms. *Phys. Rev. A* **81**, 053403 (2010).

[10] Anderson, B. M., Juzeliūnas, G., Galitski, V. M., and Spielman, I. B. Synthetic 3D Spin-Orbit Coupling. *Phys. Rev. Lett.* **108**, 235301 (2012).

[11] Lin, Y.-J., Compton, R. L., Jiménez-García, K., Porto, J. V., & Spielman, I. B. Synthetic magnetic fields for ultracold neutral atoms. *Nature* **462**, 628–632 (2009).

[12] Lin, Y.-J., Jiménez-García, K., and Spielman, I. B. Spin-orbit-coupled Bose-Einstein condensates. *Nature* **471**, 83–86 (2011). **This work demonstrated the first observation of SOC in an atomic quantum gas, and observed a quantum phase transition in the resulting two component spin-orbit coupled BECs.**

[13] Sau, J. D., Lutchyn, R. M., Tewari, S. & Sarma, Das, S. Generic New Platform for Topological Quantum Computation Using Semiconductor Heterostructures. *Phys. Rev. Lett.* **104**, (2010).

[14] Fisher, M. P. A., Weichman, P. B., Grinstein, G. & Fisher, D. S. Boson localization and the superfluid-insulator transition. *Phys. Rev. B* **40**, 546–570 (1989).

[15] Greiner, M., Mandel, O., Esslinger, T., Hansch, T. W. & Bloch, I. Quantum Phase Transition from a Superfluid to a Mott Insulator in a Gas of Ultracold Atoms. *Nature* **415**, 39–44 (2002).

[16] Radić, J., Di Ciolo, A., Sun, K. & Galitski, V. Exotic Quantum Spin Models in Spin-Orbit-Coupled Mott Insulators. *Phys. Rev. Lett.* **109**, 085303 (2012).

[17] Cole, W., Zhang, S., Paramekanti, A. & Trivedi, N. Bose-Hubbard Models with Synthetic Spin-Orbit Coupling: Mott Insulators, Spin Textures, and Superfluidity. *Phys. Rev. Lett.* **109**, 085302 (2012).

[18] M. Levin and A. Stern, Fractional topological insulators. Phys. Rev. Lett. **103**, 196803 (2009).

[19] Sedrakyan, T. A., Kamenev, A., and Glazman, L. I. Composite fermion state of spin-orbit coupled bosons. arXiv:1208.6266 (2012).

[20] Ashhab, S. and Leggett, A. J. Bose-Einstein condensation of spin-1/2 atoms with conserved total spin. *Phys. Rev. A* **68**, 063612 (2003).

[21] Cai, Z., Zhou, X. & Wu, C. Magnetic phases of bosons with synthetic spin-orbit coupling in optical lattices. *Phys. Rev. A* **85**, 061605 (2012).

[22] Stanescu, T., Anderson, B., and Galitski V., Spin-orbit coupled Bose-Einstein condensates. *Phys. Rev. A* **78**, 023616 (2008).

[23] Bychkov, Y. A. and Rashba, E. I. Oscillatory effects and the magnetic susceptibility of carriers in inversion layers. *J. Phys. C* **17**, 6039 (1984).

[24] Meier, L. *et al.* Measurement of Rashba and Dresselhaus spin--orbit magnetic fields. *Nature Physics* **3**, 650–654 (2007).

[25] Dresselhaus, G. Spin-Orbit Coupling Effects in Zinc Blende Structures. *Phys. Rev.* **100**, 580–586 (1955).

[26] Zutic, von, I., Fabian, J. & Sarma, Das, S. Spintronics: Fundamentals and applications. *Rev. Mod. Phys.* **76**, 323–410 (2004).

[27] Sinova, J., Cilcer, D., Niu, Q., Sinitsyn, N., Jungwirth, T., and MacDonald, A., Universal Intrinsic Spin Hall Effect. *Phys. Rev. Lett.* **92**, 126603 (2004).

[28] Kato, Y. K., Myers, R. C., Gossard, A. C. & Awschalom, D. D. Observation of the Spin Hall Effect in Semiconductors. Science 306, 1910–1913 (2004).

[29] Burkov, A. A. and Balents, L., Weyl Semimetal in a Topological Insulator Multilayer. Phys. Rev. Lett. **107**, 127205 (2011).

[30] Koralek, J. D., Weber, C., Orenstein, J., Bernevig, A., Zhang, S.-C., Mack, S., and Awschalom, D. Emergence of the persistent spin helix in semiconductor quantum wells. *Nature* **458**, 610-613 (2009).

[31] Kitaev, A. Y. Unpaired Majorana fermions in quantum wires. Phys.-Usp. 44, 131–136 (2001). **Proposed the idea that Majorana fermions can exist at the end of 1D superconducting wires, and idea that is directly relevant to 1D atomic Fermi gases with SOC.**

[32] Alicea, J., Oreg, Y., Refael, G., Oppen, von, F. & Fisher, M. P. A. Non-Abelian statistics and topological quantum information processing in 1D wire networks. Nat Phys **7**, 412–417 (2011).

[33] Mourik, V. *et al.* Signatures of Majorana Fermions in Hybrid Superconductor-Semiconductor Nanowire Devices. *Science*, **336**, 1003-1007 (2012).

[34] Higbie, J. & Stamper-Kurn, D. M. Periodically Dressed Bose-Einstein Condensate: A Superfluid with an Anisotropic and Variable Critical Velocity. Phys. Rev. Lett. **88**, 090401 (2002). **Proposed loading quantum degenerate gases into the laser-dressed states used in current SOC experiments.**

[35] Fu, Z., Wang, P., Chai, S., Huang, L. & Zhang, J. Bose-Einstein condensate in a light-induced vector gauge potential using 1064-nm optical-dipole-trap lasers. *Phys. Rev. A* **84**, (2011).

[36] Zhang, J.-Y. *et al.* Collective Dipole Oscillations of a Spin-Orbit Coupled Bose-Einstein Condensate. *Phys. Rev. Lett.* **109**, 115301 (2012).

[37] Wang, P. *et al.* Spin-Orbit Coupled Degenerate Fermi Gases. *Phys. Rev. Lett.* **109**, 095301 (2012). **First observation of SOC in an atomic Fermi gas.**

[38] Cheuk, L. *et al.* Spin-Injection Spectroscopy of a Spin-Orbit Coupled Fermi Gas. *Phys. Rev. Lett.* **109**, 095302 (2012). **Observation of SOC in an atomic Fermi gas, and a direct spectroscopic measurement of the SOC dispersion relation.**

[39] Campbell, D. L., Juzeliūnas, G. & Spielman, I. B. Realistic Rashba and Dresselhaus spin-orbit coupling for neutral atoms. *Phys. Rev. A* **84**, (2011).

[40] Wang, C., Gao, C., Jian, C.-M. & Zhai, H. Spin-Orbit Coupled Spinor Bose-Einstein Condensates. *Phys. Rev. Lett.* **105**, 160403 (2010).

[41] Ho, T.-L. & Zhang, S. Bose-Einstein Condensates with Spin-Orbit Interaction. *Phys. Rev. Lett.* **107**, 150403 (2011).

[42] Wu, C.-J., Mondragon-Shem, I. & Zhou, X.-F. Unconventional Bose—Einstein Condensations from Spin-Orbit Coupling. *Chinese Phys. Lett.* **28**, 097102 (2011).

[43] Giorgini, S., Pitaevskii, L. P. & Stringari, S. Theory of ultracold atomic Fermi gases. Rev. Mod. Phys. **80**, 1215–1274 (2008).

[44] Ketterle, W. & Zwierlein, M. W. Making, probing and understanding ultracold Fermi gases. Proc. Int. Schl. Phys., Varenna, edited by M. Inguscio, W. Ketterle, and C. Salomon (IOS Press, Amsterdam) 2008.

[45] Magarill, L. I. and Chaplik, A. V. Bound States in a Two-Dimensional Short Range Potential Induced by the Spin-Orbit Interaction. *Phys. Rev. Lett.* **96**, 126402 (2006).

[46] Gong, M., Tewari, S., and Zhang, C., BCS-BEC Crossover and Topological Phase Transition in 3D Spin-Orbit Coupled Degenerate Fermi Gases. *Phys. Rev. Lett.* **107**, 195303 (2011).

[47] Yu, Z.-Q. and Zhai, H., Spin-Orbit Coupled Fermi Gases across a Feshbach Resonance. *Phys. Rev. Lett.* **107**, 195305 (2011).

[48] Veillette, M., Sheehy, D., Radzihovsky, L. & Gurarie, V. Superfluid Transition in a Rotating Fermi Gas with Resonant Interactions. *Phys. Rev. Lett.* **97**, 250401 (2006).

[49] Levinsen, J., Cooper, N. R., and Gurarie, V. Strongly Resonant $p$-Wave Superfluids. *Phys. Rev. Lett.* **99**, 210402 (2007).

[50] Regal, C. A., Ticknor, C., Bohn, J. L. & Jin, D. S. Tuning $p$-Wave Interactions in an Ultracold Fermi Gas of Atoms. *Phys. Rev. Lett.* **90**, 053201 (2003).

[51] Williams, R. A. *et al.* Synthetic Partial Waves in Ultracold Atomic Collisions. Science **335**, 314–317 (2012).

[52] Zhang, C., Tewari, S., Lutchyn, R. M. & Sarma, Das, S. $p_x+ip_y$ Superfluid from $s$-Wave Interactions of Fermionic Cold Atoms. *Phys. Rev. Lett.* **101**, 160401 (2008).

[53] Massignan, P., Sanpera, A. & Lewenstein, M. Creating p-wave superfluids and topological excitations in optical lattices. *Phys. Rev. A* **81**, 031607 (2010).

[54] Seo, K., Han, L. & Sá de Melo, C. Emergence of Majorana and Dirac Particles in Ultracold Fermions via Tunable Interactions, Spin-Orbit Effects, and Zeeman Fields. *Phys. Rev. Lett.* **109**, 105303 (2012).

[55] Schnyder, A. P., Ryu, S., Furusaki, A. & Ludwig, A. W. W. Classification of topological insulators and superconductors in three spatial dimensions. Phys. Rev. B **78**, 195125 (2008).

[56] A. Kitaev, Periodic table for topological insulators and superconductors. *Advances in theoretical physics, AIP conference proceedings*, **1134** (2009).

[57] Moore, J. E., and Balents, L., Topological invariants of time-reversal-invariant band structures. *Phys. Rev. B* **75**, 121306(R) (2007).

[58] Gell-Mann, M. Symmetries of Baryons and Mesons. *Phys. Rev.* **125**, 1067–1084 (1962).

[59] Lindner, N. H., Refael, G., and Galitski, V., Floquet Topological Insulator in Semiconductor Quantum Wells. *Nature Physics* **7**, 490--495 (2011).

[60] Jiang, L., *et al.* Majorana Fermions in Equilibrium and Driven Cold Atom Quantum Wires. *Phys. Rev. Lett.* **106**, 220402 (2011).

[61] Shapere, A. and Wilczek, F., *Geometric Phases in Physics* (World Pacific, Singapore 1989).

[62] Dalibard, J., Gerbier, F., Juzeliūnas, G. & Öhberg, P. Artificial gauge potentials for neutral atoms. *Rev. Mod. Phys.* **83**, 1523–1543 (2011).

[63] Jaksch, D. & Zoller, P. Creation of effective magnetic fields in optical lattices: the Hofstadter butterfly for cold neutral atoms. *New Journal of Physics* **5**, 56 (2003).

[64] Aidelsburger, M. *et al.* Experimental Realization of Strong Effective Magnetic Fields in an Optical Lattice. *Phys. Rev. Lett.* **107**, 255301 (2011).

[65] Struck, J. *et al.* Tunable Gauge Potential for Neutral and Spinless Particles in Driven Optical Lattices. *Phys. Rev. Lett.* **108**, (2012).

[66] Zhu, S.-L., Fu, H., Wu, C. J., Zhang, S. C. & Duan, L. M. Spin Hall Effects for Cold Atoms in a Light-Induced Gauge Potential. *Phys. Rev. Lett.* **97**, 240401 (2006).

[67] Bermudez, A. et al. Wilson Fermions and Axion Electrodynamics in Optical Lattices. Phys. Rev. Lett. 105, 190404 (2010).

[68] Zohar, E., Cirac, J. & Reznik, B. Simulating Compact Quantum Electrodynamics with Ultracold Atoms: Probing Confinement and Nonperturbative Effects. *Phys. Rev. Lett.* **109**, 125302 (2012).

[69] Banerjee, D. *et al.* Atomic Quantum Simulation of Dynamical Gauge Fields Coupled to Fermionic Matter: From String Breaking to Evolution after a Quench. Phys. Rev. Lett. **109**, 175302 (2012).